\documentclass[aps,prb,twocolumn,showpacs,floatfix,superscriptaddress,amssymb,amsmath]{revtex4}
\usepackage{graphicx}
\usepackage{amssymb}

\newcommand{\be}{\begin{equation}}
\newcommand{\ee}{\end{equation}}
\newcommand{\bea}{\begin{eqnarray}}
\newcommand{\eea}{\end{eqnarray}}

\newcommand{\vare}{\varepsilon}

\newcommand{\re}{\mbox{e}}
\newcommand{\ba}{\begin{array}}
\newcommand{\ea}{\end{array}}
\def\nn{\nonumber\\}
\def\vare{{\varepsilon}}
\def\up{\uparrow}
\def\down{\downarrow}

\begin{document}


\title{Quantum spin Hall phase in neutral zigzag graphene ribbons}
  \author{Mahdi Zarea}
  \author{and Nancy Sandler}
\address{Department of Physics and Astronomy, Nanoscale and Quantum Phenomena Institute, and 
Condensed Matter and Surface Science Program,\\Ohio University, Athens, OH 45701-2979}

\begin{abstract}
We present a detailed description of the nature of the wavefunction and spin distribution of the zero energy modes of zigzag graphene ribbons ($ZGR$s) in the presence of the intrinsic spin-orbit (I-SO) interaction. These states characterize the quantum spin Hall (QSH) phase in graphene ribbons. We provide analytic expressions for wavefunctions and show how these evolve as the strength of the interaction and the ribbon width are changed. For odd-width ribbons, we show that its insulating nature precludes the existence of a QSH phase. For these systems the I-SO interaction is predicted to have a stronger effect as shown by the enhancement of the gap as the interaction strength is turned on.
\end{abstract}


\maketitle
\section{Introduction}

Much of the interest in graphene, the mono-layer of carbon atoms arranged in a honeycomb lattice obtained for the first time few years ago\cite{geim1,kim1} lies on its potential applications in electronic circuitry. As methods of fabrication improve, samples with higher mobilities and better conducting properties are produced, as shown by the recent experiments where mobilities of the order of $20 \times 10^4 cm^2/Vs$ were obtained\cite{kim2} . It is thus reasonable to expect  that controlled design of sample sizes as well as of edge terminations will be possible in a near future. Of particular interest for device applications are graphene ribbons or wires where the role of confinement has greater influence  on transport properties. Several theoretical studies \cite{fujita1,nakada1,fertig1,fertig2,antonio1} have focused on various properties of graphene ribbons  modeled with two different edge terminations -and further combinations of them to represent more complicated edges. These standard model terminations are known as armchair and zigzag edges. While the physics of armchair ribbons shows the phenomena expected from confinement of a graphene sheet (the band-structure of the two-dimensional material is fully reproduced in the limit of an infinite wide ribbon)\cite{fujita1}, zigzag ribbons possess zero energy modes that are highly localized along the edges of the sample. These modes, with a topological origin, play an important role on the magnetic properties of the ribbon and have been the subject of extensive analytic and numerical research\cite{complete}. 
In this paper we present a detailed analysis of the properties of zigzag graphene ribbons (ZGRs) that includes the effects of the intrinsic spin-orbit (I-SO) interaction introduced by Kane and Mele\cite{kane1,kane2} for graphene. As predicted in that work, the I-SO interaction gives origin to the quantum spin Hall (QSH) phase in two-dimensional graphene, a new state of matter characterized by an insulating bulk and zero energy chiral modes that are fully spin polarized. The analytic solution of a tight-binding model for the ribbon allows to write explicit expressions for the wavefunctions as functions of the I-SO interaction strength and the ribbon width. We also are able to show that only even-width ribbons can exhibit the QSH phase and show how the
gap in the energy spectrum of odd-width ribbons changes with the I-SO coupling.  

\section{Model of graphene with I-SO interactions}

\begin{figure}
\includegraphics[width=6cm]{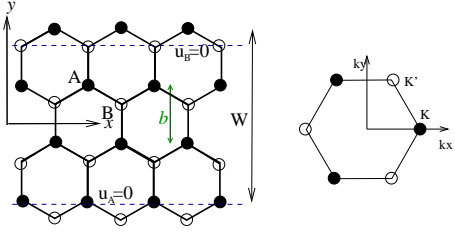}
\caption{The left panel shows a 2d graphene lattice with the sublattice sites labeled as  $A$ and $B$ and the zigzag ribbon boundary conditions (see text). The right panel shows the first Brillouin zone of graphene with the two Dirac points chosen at $K=({4\pi\over3a},0)$ and $K'=({2\pi\over3a},{\pi\over b})$.} 
\label{Fig.1}
\end{figure}

The crystalline structure of 2D graphene is described in terms of two sublattices $A$ and $B$ with a lattice constant $a=2.4~\AA$ (see Fig.~(\ref{Fig.1})). The corresponding two-valued wavefunctions are spinors whose components describe the occupation of each sublattice.  
A nearest neighbor tight-binding model with SU(2) spin symmetry produces a Hamiltonian matrix and corresponding wavefunctions (for a given spin) of the form:  
\be
H  =  \left( \begin{array}{cc}
             0 & \phi  \\
             \bar\phi & 0
             \end{array}\right)
~~~~%
\Psi = \left(\begin{array}{c} 
    u_{A}         \\
    u_{B}        
             \end{array}\right)
\label{hamiltonian}
\ee
with $\phi(k_x,k_y) =  t( \re^{ik_y2b/3}+2\cos{\frac{k_xa}{2}}\re^{-ik_yb/3})$  and  $b=a\sqrt{3}/2$. In these expressions $\bar\phi$ is given by $\bar\phi(k_x,k_y)=\phi(k_x,-k_y)$ and for real values of $k_y$, $\bar\phi = \phi^*$. 

For convenience we introduce a global gauge transformation  $u_B\to u_B\re^{ik_yb/3}$ which correspond to a redefinition of $\phi$ as 
\be
\phi(k_x,k_y) =  t( \re^{ik_yb}+2\cos{\frac{k_xa}{2}}).
\ee
The transformation amounts to label all the atoms along each 
zigzag line by a unique $y$-coordinate.

The eigenvalues of (\ref{hamiltonian}) are obtained in a straightforward manner as $E=\pm\vare=\pm\sqrt{\phi\bar\phi}$ 
and the corresponding eigenvectors are given by:
\be
\Psi_{\pm}=\left( \begin{array}{c}
 \re^{i\alpha/2}  \\
 \pm \re^{-i\alpha/2}
\end{array}\right)\re^{ik_xx}\re^{ik_yy}
\ee
where $\alpha$ is defined by $\phi=\vare\re^{i\alpha}$. For neutral graphene, 
 $\Psi_+$ ($\Psi_-$) represents solutions with $E > 0$  $(E < 0)$ and 
refers to electron (hole) conduction (valence) bands. In this language,
particle-hole symmetry implies that for each electron state with 
energy $E=\vare$ and eigenstate $\alpha$, there is a hole state with $E=-\vare$ and eigenstate given by $\alpha+\pi$. At the Dirac points
(two independent degeneracy points in the Brillouin zone) $\phi=0$ or $\bar\phi=0$.

To include the I-SO interaction we use its real space representation  in second quantization that involves a spin-dependent second-neighbor hopping term:  $H_{I-SO} \sim i t' (\mathbf{d}_{ik} \times \mathbf{d}_{kj})_z c^{\dagger}_is^zc_j$, with $\mathbf{d}_{ik}$ as first neighbor lattice vectors connecting electrons at  positions $i$ and $k$, $c^{\dagger}_{i}$ the electron creation operator at site $i$ and ${s^z}$ the spin along the $z$-direction (we have chosen the $z$-direction perpendicular to the graphene plane for convience) \cite{kane1,mehdi1}. Notice that this expression satisfies all 
the symmetries of the graphene lattice\cite{manes}. Reported values for the interaction strength $t'$ obtained with various ab-initio calculations range from $1.2K$ to $10mK$\cite{trickey}.

\begin{figure}
\includegraphics[width=.2\textwidth]{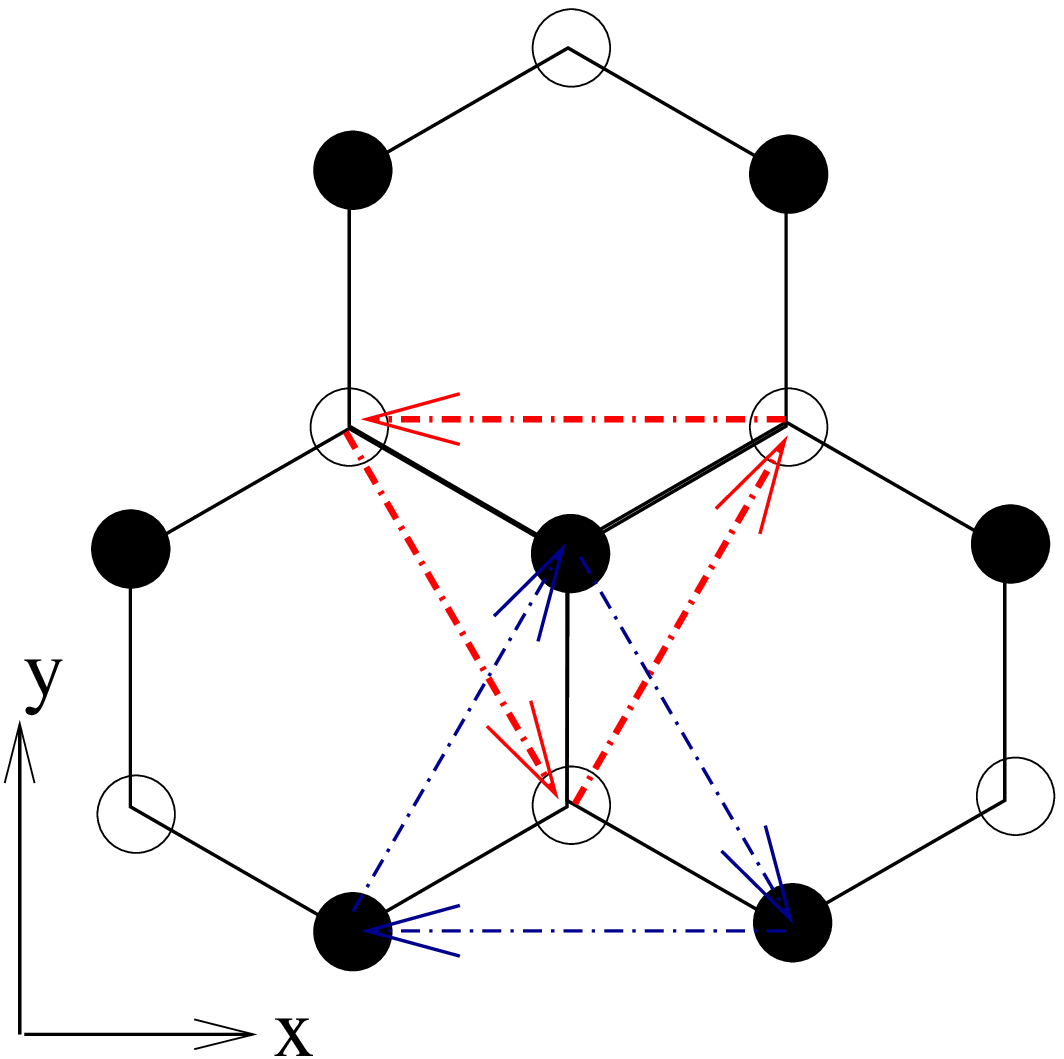}
\includegraphics[width=.2\textwidth]{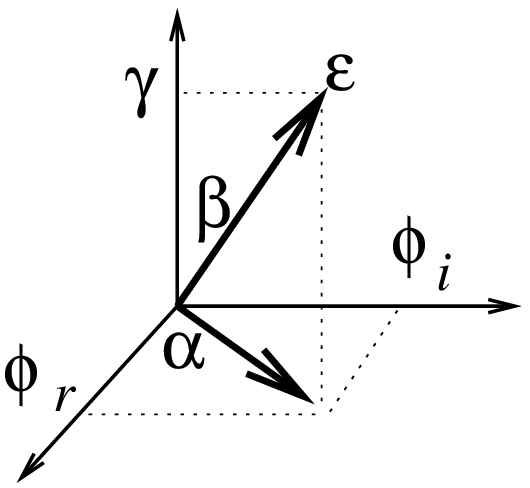}
\caption{Left: I-SO term with second-neighbor hopping of an electron in 
sub-lattice $A$, coupled to a spin on a $B$ site. The I-SO term
 results from adding clock- and anticlockwise motions and exchanging the role of the two sublattices. Right: Definitions of $\alpha$ and $\beta$. 
}
\label{Fig.23}
\end{figure}

The total Hamiltonian in reciprocal space reads:
\be
H  =  \left( \begin{array}{cc}
  s\gamma & \phi  \\
  \bar\phi & -s\gamma  \end{array}\right)
\label{eq:hamiltonian}
\ee
where $\gamma(k_x,k_y) =  2t'(\sin k_xa-2\sin{\frac{k_xa}{2}}\cos{k_yb})$
and $s=\pm$ stands for spin-up/down electron. The effect of the I-SO interaction as it appears in Eq.(\ref{eq:hamiltonian}) is to introduce 
opposite staggered magnetic fields acting on opposite spins. Fig.(\ref{Fig.23}) shows a schematic representation of one possible process introduced by the I-SO coupling. Below we consider only spin-up electrons. Notice that to obtain the corresponding expression for spin-down electrons it is just enough to make the replacement $t'\to -t'$.  

Spin-up electrons and holes in states $k_x, k_y$ have energies given by $E=\pm\vare=\pm\sqrt{\phi\bar\phi+\gamma^2}$. The main effect of the I-SO interaction is to open a gap at the Dirac points with the system becoming a bulk-insulator\cite{kane1}.
 
As an example, electrons eigenstates  ($E>0$) are given by:
$\Psi_+=\chi_{k_y}\re^{ik_yy}\re^{ik_xx}$ where
\bea
\chi_{k_y}&=&\left( \begin{array}{c}
  \cos{\beta\over2}\re^{i\alpha/2}  \\
  \sin{\beta\over2}\re^{-i\alpha/2} \end{array}\right)
\label{eigen-bulk}
\eea
and the angles $\alpha, \beta$ are defined by $\gamma=\vare\cos{\beta}$ and
$\phi=|\phi|\re^{i\alpha}$ . The corresponding hole state is 
$\Psi_{-}=\eta_{k_y}\re^{ik_yy}\re^{ik_xx}$ with $\eta$  obtained from $\chi$ by the replacement $\beta\to \pi-\beta$.

\section{ Zigzag graphene ribbons}

\begin{figure}
 \includegraphics[width=.5\textwidth]{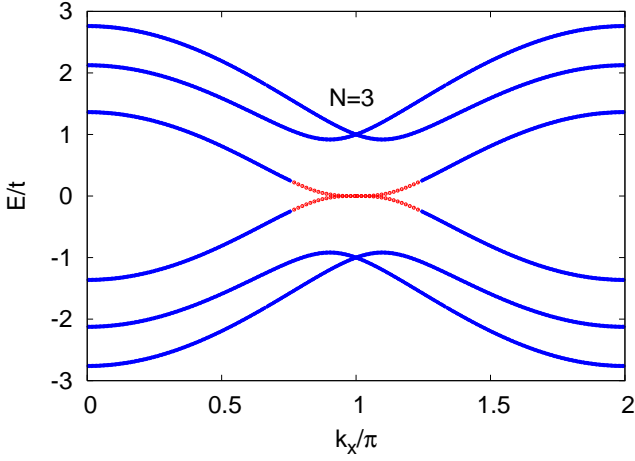}
 \caption{Energy bands of a zigzag ribbon with $W=4b$ 
in the absence of the I-SO interaction. Each band is doubly 
degenerate due to the spin SU(2) symmetry. The zero energy (edge) band (red line) is obtained when the wavenumber $k$  in (\ref{zgr-cnd}) takes an imaginary value. The edge state starts at the point $k_x^0$ where
$\cos(k_x^0a/2)= \pm {N\over 2N-2}$}
 \label{Fig.4}
\end{figure}

To study the role of the I-SO interaction in a confined geometry we analyze a
zigzag graphene  nanoribbon, defined according to Fig.~(\ref{Fig.1}).
Hard-wall boundary conditions are imposed by setting $u_A=0$  on the lower border and $u_B=0$ on the upper border \cite{fujita1,nakada1,fertig1,hikihara1}.
For symmetry reasons we chose the origin of the $y$-axis in the center of the ribbon and the boundary conditions become:
\be
u_A(y=-W/2)=0, ~~u_B(y=W/2)=0. 
\label{dbc}
\ee
where $W$ is the ribbon width with a number $N = W/b-1$ of chains inside it.
The wavefunction of the ZGR is found as follows:
Because of translation invariance along the $x$-axis, $k_x$ is a good quantum number. For a given $k_x$, the total wavefunction must be a superposition of degenerate states with different $k_y$ values. In the absence of I-SO there are only two degenerate spinors for a fixed value of $k_x$: $k_y = k$ and $k_y = -k$. Therefore the wavefunction is the superposition of these
two spinors: $\Psi=a\Psi(k_x, k)+b\Psi(k_x, -k)$. Application of
the boundary conditions given in Eq.(\ref{dbc}), renders $b=-a$ with
\bea
\Psi&=&C\left( \begin{array}{c}
 \sin(\alpha/2+ky-n\pi/2)  \\
 \sin(-\alpha/2+ky-n\pi/2)\end{array}\right)\re^{ik_xx}
\eea
where $k$  satisfies
\be
\alpha-kW=n\pi \label{zgr-cnd}
\ee
and $C={1\over\sqrt{2L}}\sqrt{|\Re{k}/\sinh(\Re{k})-\Im{k}/\sin(\Im{k}|}$ is the normalization factor. Notice that a peculiar feature of a ZGR is that $k$ can take complex values between two Dirac points \cite{fujita1,nakada1,fertig1}. Fig.~(\ref{Fig.4}) shows the conduction bands of a ribbon with $W=4b$.

The expression of an edge state wavefunction with $k_xa>\pi$ is given by
\bea
\Psi&=&C\left( \begin{array}{c}
  \sinh(q(y+W/2))  \\
  -\sinh(q(y-W/2))\end{array}\right)\re^{ik_xx}
\eea
where $k_y=-iq$ is purely imaginary and satisfies Eq.(\ref{zgr-cnd}) with
$n=0$ and $\alpha=\pi+i\alpha_0$ and $\alpha_0=-qW$. For $k_xa<\pi$ the edge state wavefunction is given by
\bea
\chi&=&C\left( \begin{array}{c}
 \sinh(q+i\pi)(y+W/2)  \\
 \sinh[(q+i\pi)(y-W/2)+i\pi N]\end{array}\right)\re^{ik_xx} \label{edge-kgp}
\eea
where $k_y=-\pi/b-iq$  and satisfies Eq.(\ref{zgr-cnd}) with
$n=N$ and $\alpha=i\alpha_0$ and $\alpha_0=-qW$. 

\subsection{Zigzag ribbons with I-SO interactions}

\begin{figure}
 \includegraphics[width=.5\textwidth]{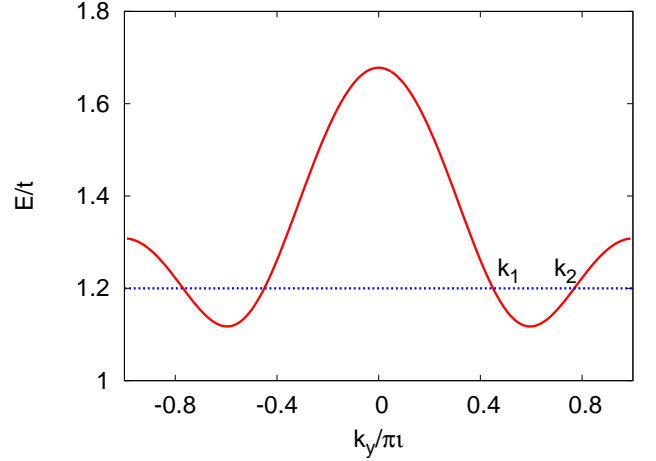}
 \caption{In the absence of I-SO, for a fixed value of $k_x$,
 there are only two degenerate states at $k_y=\pm k$. In
 the presence of the I-SO interaction there are
four degenerate states at $k_y=\pm k_1,\pm k_2$.  }
 \label{Fig.5}
\end{figure}

Because the I-SO interaction involves second-neighbor hoping there is a new set of boundary conditions that has to be satisfied: besides those given in Eq.(\ref{dbc}) it is necessary to impose two extra conditions given by $u_A(y = W/2)=u_B(y = -W/2)=0$. Correspondingly, for a fixed value of $k_x$, there are now four degenerate states at $k_y=(\pm k_1;\pm k_2)$ as shown in Fig.\ref{Fig.5}. 

Note that for some energies and values of $k_x$, the numbers $k_1$ and $k_2$ can be complex. To guarantee the degeneracy of these points, the energy condition $E(\pm k_1)=E(\pm k_2)$ is transformed into the condition
\be
\cos k_1b+\cos k_2b=2c(1-\tau^2)
\label{deg}
\ee
with  $\tau^2=1/(8t'^2\sin^2 {\frac{k_xa}{2}})$ and $c = \cos(k_xa/2)$.
 
As an example let's consider the wavefunction for a spin-up state with $k_x a > \pi$ given by the linear combination of the four degenerate spinors:
\bea
\Psi_{R\up} = a_1\chi_{k_1}\re^{ik_1y}-b_1\chi_{-k_1}\re^{-ik_1y}&&\nn
+a_2\chi_{k_2}\re^{ik_2y}-b_2\chi_{-k_2}\re^{-ik_2y}. &&
\label{main-wf}
\eea
Nontrivial solutions of Eq.(\ref{main-wf}) exist only if the 
condition 
\bea
\big((\gamma_1-\gamma_2)^2+(\phi_1-\phi_2)(\bar\phi_1-\bar\phi_2)\big)
\sin^2(qW/2)&&   \nn
=\big((\gamma_1-\gamma_2)^2+(\phi_1-\bar\phi_2)(\bar\phi_1-\phi_2)\big)
\sin^2(kW/2) &&
\label{det-c}
\eea
is satisfied. In these expressions $\gamma_{j}=\gamma(k_{y}=k_{j})$, 
and $k=(k_{1} + k_{2})/2$, $q=(k_1-k_2)/2$. 
The two conditions given by Eqs.(\ref{deg}) and (\ref{det-c}) uniquely determine $k_1,k_2$ as functions of $t,t'$ and $W$. The effects of the I-SO interaction on the band structure have been described in a previous work \cite{mehdi2}. There it was shown that the quasi-degeneracy of the zero energy mode is lifted and the dispersion around $k_x = \pi$ becomes linear. This can be shown to be the exact solution in the limit of a semi-infinite ribbon, i.e, when $t'/t\gg b/W$. In this case, a simplified expression for the edge-band dispersion can be found because the conditions imposed by Eqs. (\ref{deg}) and (\ref{det-c}) are simplified as follows:
$\cosh2qb=1+\tau^2$ and $\cosh2kb=4c^2(1-\tau^2)^2/(2+\tau^2)-1$. The dispersion relation of the edge sates becomes:
 \be
\vare=\pm6t'\sin(k_xa)\sqrt{1+16t'^2\sin^2(k_xa/2)}\label{energy-inf}
\ee 
which near $k_xa=\pi$ can be approximated by $\vare\approx\pm6t'k_xa$ with 
the velocity of right (left) movers given by $\hbar v=\pm 6t'a$. 
\begin{figure}
\includegraphics[width=.45\textwidth]{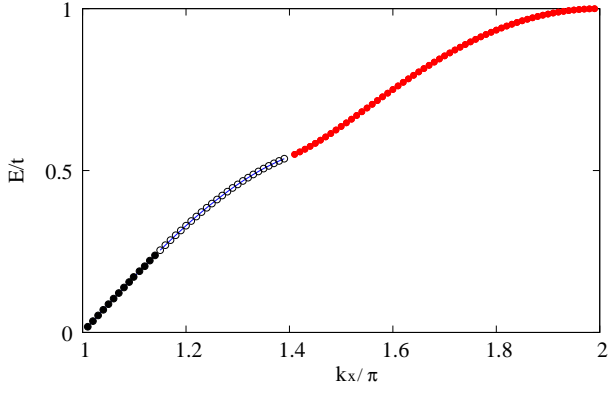}
  \caption{Lowest energy band of a semi-infinite zigzag graphene ribbon with a I-SO coupling constant $t'=.1t$. The edge state dispersion is given exactly by Eq. (\ref{energy-inf}). Different colors represent different values of $k_y$ as described in Ref.\cite{mehdi2}.}
\label{Fig.6}
\end{figure}

\begin{figure}
      \includegraphics[width=.45\textwidth]{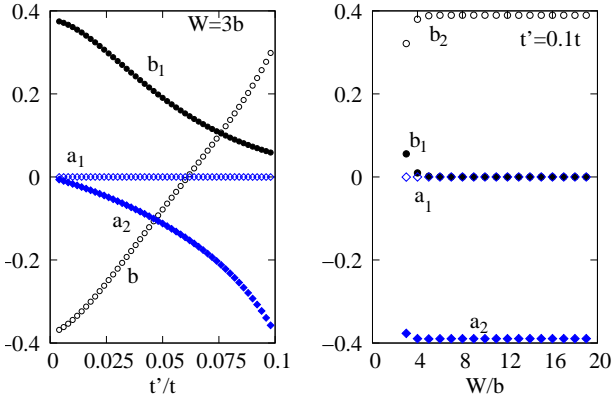}
	\caption{Coefficients of the edge state wavefunction as a function of the I-SO coupling $t'$ for $W=3b$ (left panel) and as a function of the ribbon width $W$ for $t'=.1t$ (right panel).}
	\label{Fig.7}
      \end{figure}

To gain better understanding of the properties of the wavefunction associated with this mode, we analyze the behavior of the coefficients appearing in Eq.(\ref{main-wf}) as functions of the I-SO coupling strength and the ribbon width. The expressions for these coefficients are:
\bea
&&a_1=<\eta_{-k_2}|\chi_{-k_1}><\eta_{-k_1}|\chi_{k_2}>\sinh kW\sinh k_2W\nn
&&b_1=<\eta_{-k_2}|\chi_{k_1}><\eta_{-k_1}|\chi_{k_2}>\sinh qW\sinh k_2W\nn
&&a_2=<\eta_{-k_2}|\chi_{-k_1}><\eta_{-k_1}|\chi_{k_1}>\sinh kW\sinh kW\nn
&&b_2=<\eta_{-k_1}|\chi_{k_2}><\eta_{-k_1}|\chi_{k_1}>\sinh kW\sinh qW
\eea
Notice that in the limit $t' \to 0$, $\re^{ik_1b}\approx-2c\tau^2$, $b_1\to-b_2$ while  $a1<<a_2$ and both of them go to zero as $t'\to 0$ as shown in Fig.\ref{Fig.7}. In this way, the wavefunction of a ZGR in the absence of I-SO interaction is recovered. On the other hand for a semi-infinite ribbon: $a_1$ and $b_1 \to 0$ while $a_2 \to -b_2$ and remains finite.
      
Further insight into the effect of the I-SO interaction is gained when the spin distribution of the edge state is studied. Let us take for example spin-up states with $k_x a > \pi$ with a given state in the conduction band (electron or right mover state) represented by 
\bea
\chi_{R\up}=\left(\begin{array}{c}
  \varphi_A(y)  \\
  \varphi_B(y)
\end{array}\right)&&
\eea

The corresponding valence band state (hole or left mover state) is given by
\bea
\chi_{L\up}=\left(\begin{array}{c}
  -\varphi_B(-y)  \\
  \varphi_A(-y)
\end{array}\right).&&
\eea

\begin{figure}
  \includegraphics[width=.5\textwidth]{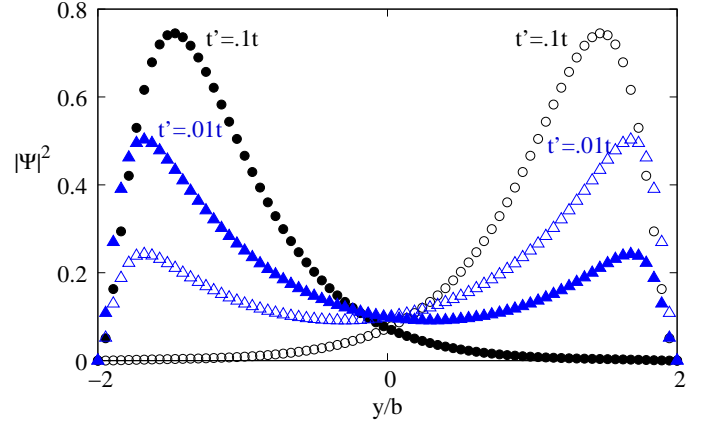}
  \caption{Probability distribution for spin-up right- (filled symbols)
    and left- mover (empty symbols) as a function of the position across the ribbon, for I-SO coupling strengths $t' = 0.1t$ (circles) and $t' = 0.01t$ (triangles). The ribbon width is $W = 4b$.
  }
  \label{Fig.8}
\end{figure}

The spin-down electron states can be obtained by the replacement $t'\to-t'$. The corresponding wave functions are:
\bea
\chi_{R\down}=\left(\begin{array}{c}
  \varphi_B(-y)  \\
  \varphi_A(-y)
\end{array}\right)
\eea
for a right mover and
\bea
\chi_{L\down}=\left(\begin{array}{c}
  -\varphi_A(y)  \\
  \varphi_B(y)
\end{array}\right)&&
\eea
for a left mover. In Fig.(\ref{Fig.8}) we plot the probability distribution for a 
spin-up right (spin-down left) mover as a function of the 
position $y$ across the ribbon for two values of the I-SO coupling strength.  Notice that the predicted localized\cite{kane1} wavefunction becomes spin-polarized as as the I-SO interaction is turned on.

\section{Ribbon's width and I-SO interaction}

\begin{figure}
 \includegraphics[width=.5\textwidth]{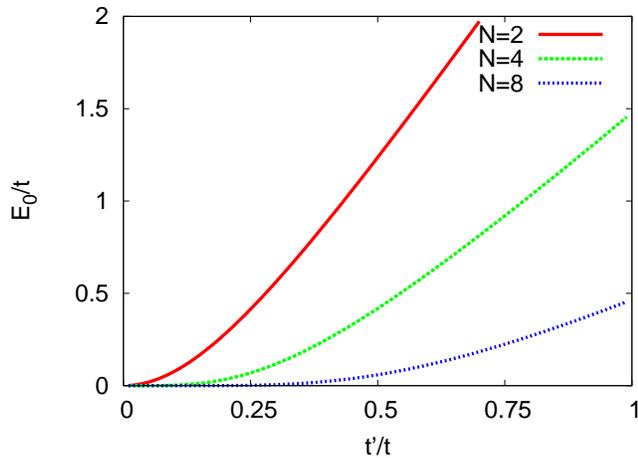}
 \caption{Band gap $\Delta=2E_0$ for odd width zigzag ribbons (N even) at $k_xa=\pi$ as a function of the I-SO coupling $t'$ for $N=2,4,8$.  }
 \label{Fig.9}
\end{figure}

The role of the I-SO interaction becomes more pronounced for certain widths of zigzag ribbons. Ribbons with even widths $W=2M b$ (where M is an integer) have an odd number N of chains inside and are metals in which edge states from conduction and valence bands become degenerate at $k_xa=\pi$. At the same time, ribbons with odd widths  $W=(2M+1) b$ (with an even number N of chains inside) are insulators \cite{mehdi2}. Thus, it is expected that the effects of the I-SO interaction will be more evident for odd-width ribbons by increasing the size of the gap. 
In fact, from Eq. (\ref{deg}) it is possible to show that at $k_x a = \pi$:
\be
k_yb=\pi+iqb~~\sinh(qb)=t/4t'
\ee
Furthermore, for  $N$ even  Eq.(\ref{det-c}) reads
\be
\sinh^2q=\tau^2(\cosh2qW-\cosh2q)/2(1+\cosh2qW)
\ee 
which results in the following energy for the edge state at $k_xa=\pi$
\be
E_0=\pm t\cosh(k_yb)/\cosh(k_yW)\label{bandgap}.
\ee
For small $t'$ the energy of this mode increases and the gap ($\Delta = 2 E_0$) between conducting and valence bands scales
as $\Delta\approx t(2t'/t)^N$. In Fig.\ref{Fig.9} we plot a typical dependence of the energy $E_0$ as a function of $t'$ for several values of $N$.

\section{Conclusions}

We have presented a detailed analytical description of the effects of the intrinsic spin-orbit (I-SO) interaction on the wavefunction and spin-distribution properties of zigzag graphene nanoribbons. Hard-wall boundary conditions in the presence of the I-SO interaction introduce a new set of 
equations to be satisfied that affect both sublattices. Thus, to obtain the wavefunction for the zero energy mode it is necessary to mix four different spinors in contrast with the two needed in absence of the I-SO interaction. An analysis of the role played by each spinor (at fixed values of the coupling strength) as the ribbon width is changed shows how the wavefunction in the semi-infinite ribbon limit is recovered. These results allows to track back the origin of the predicted spin-polarized
chiral edge states. We also provide explicit expressions for the corresponding wavefunctions that fully describe the spin probability distribution and show an increasing spin polarization as the interaction strength increases. 
The effects are expected to be more immediate for odd-width zigzag ribbons where the I-SO interaction is predicted to enhance a nascent gap that makes these systems insulators.

\section{Acknowledgments}

We acknowledge discussions with C. B\"{u}sser and S.E. Ulloa. This work was partially supported by the Ohio University Postdoctoral Fellowship program and by the National Science Foundation under grants NSF-DMR 0710581 and NSF PHY05-51164.


\begin{thebibliography}{00}

\bibitem{geim1}K.S. Novoselov {\it et. al.}, Nature {\bf 438}, 197-200 (2005).
\bibitem{kim1}Y. Zhang {\it et. al.}, Nature {\bf 438}, 201 (2005).
\bibitem{kim2}Bolotin KI, Sikes KJ, Hone J, et al., Phys. Rev. Lett. {\bf 101}, 096802, (2008).
\bibitem{fujita1}M. Fujita, J. Phys. Soc. Jap. {\bf 65}, 1920 (1996).
\bibitem{nakada1}K. Nakada, M. Fujita, G.Dresselhaus and M.S. Dresselhaus, Phys. Rev. B {\bf 54}, 17954 (1996).
\bibitem{fertig1}L. Brey and H.A Fertig, Phys. Rev. B {\bf 73}, 235411 (2006).
\bibitem{fertig2}L. Brey and H.A Fertig, Phys. Rev. B {\bf 75}, 125434 (2007).
\bibitem{antonio1}See for instance A. H. Castro Neto {\it et. al.}, Rev. Mod. Phys. {\bf 81}, 109 (2009).
\bibitem{complete}Y.W. Son, M.L. Cohen and S. G. Louie, Nature {\bf 444}, 347-349 (2006) and references therein; S.Cho, Y.F. Chen and M.S. Fhurer, cond-mat/0706.1597; L. Tapaszto et. al., cond-mat/0806.1662; Y.W. Son, M.L. Cohen and S.G. Louie, Phys. Rev. Lett. {\bf 97}, 216803 (2006); L.Yang et al., Phys. Rev. Lett. {\bf 99}, 186801 (2007);  Wu {\it et al}, PRL {\bf 96}, 106401 (2006); L. Pisani et al., Phys. Rev. B {\bf 75}, 064418 (2007).
\bibitem{kane1}C.L. Kane and E.J.  Mele, Phys. Rev. Lett. {\bf 95}, 226801 (2005)
\bibitem{kane2}C.L. Kane and E.J.  Mele, Phys. Rev. Lett. {\bf 95}, 146802, (2005).
\bibitem{mehdi1}M. Zarea and N. Sandler, Phys. Rev. Lett. {\bf 99}, 256804 (2007).
\bibitem{manes}J.L. Manes, F. Guinea and M.A.H. Vozmediano, Phys. Rev. B {\bf 75}, 155424 (2007).
\bibitem{trickey}J.Boettger and S.Trickey Phys. Rev. B {\bf 75}, 121402(R) (2007);   H. Min et. al.  Phys. Rev. B {\bf 74}, 165310 (2006); D. Huertas-Hernando,   F. Guinea  and Arne Brataas,Phys. Rew.B {\bf 74},155426 (2006).
\bibitem{hikihara1}T. Hikihara, X. Hu, H.-H. Lin and C.-Y. Mou, Phys. Rev. B {\bf 68}, 035432 (2003).
\bibitem{mehdi2}M. Zarea, C. B{\"u}sser and N. Sandler, Phys. Rev. Lett. {\bf 101}, 196804 (2008).

\end{thebibliography}
\end{document}